\documentclass[prb,aps,twocolumn,showpacs]{revtex4-1}
\usepackage{amsmath}
\usepackage{graphicx}
\usepackage{epsfig}
\usepackage{amssymb}
\usepackage{color}
\usepackage{subfig}
\DeclareMathAlphabet{\mathpzc}{OT1}{pzc}{m}{it}

\begin{document}
\title{Resonant Bragg quantum wells in hybrid photonic crystals}
\author{A. D'Andrea and N. Tomassini}
\affiliation{Istituto dei Sistemi Complessi, CNR, C.P. 10, 
Monterotondo Stazione, Roma I-00015}
\date{\today }
%
%
\begin{abstract}
The exciton-polariton propagation in resonant hybrid (isotropic/anisotropic) periodic stacks, with misaligned in plane anisotropy and  Bragg photon frequency in resonance with Wannier exciton of 2D quantum wells, is studied by self-consistent theory and in the effective mass approximation. The optical tailoring of this new class of resonant Bragg reflectors, where the structural periodicity of a multi-layer drives the  periodicity of the in-plane optical $\hat C$ axis orientation, is computed for symmetric and non-symmetric elementary cell by conserving strong radiation-matter coupling and photonic band gap. We will demonstrate, by selected numerical examples, that the behavior of the so called intermediate dispersion curves (IDC), that drop between upper and lower branches of the lowest energy band gap, are strongly dependent from in-plane $\hat C$ axis orientation.  Therefore, we guess that this class of hybrid meta-materials is promising  for  new trapping light optical devices based on IDC behavior.
\end{abstract}
\maketitle
%
%
\section{INTRODUCTION}
The manipulation of the optical properties in order to obtain coherent radiative coupling among a collection of emissive species was proposed by R.H. Dicke\cite{1}  in 1954. The optical behavior of the composite exciton-polariton, propagating in a resonant 1D Bragg quantum wells, was studied by E.L. Ivechenko et all. \cite{2} in 1994, while the saturation effect, that transforms the super-radiant mode into a polaritonic Bloch band, usually obtained for rather large quantum well number ($N \ge 200$), was theoretically investigated by K. Cho et all. \cite{3,4}  in 2002 and measured by J.P. Prineas at all.\cite{5}  in the same year.
 
It is well known  that Wannier exciton is intrinsically a giant dipole excitation, since the microscopic dipole transitions of the elementary cells of a semiconductor are coherently coupled in a volume determined by its Bohr radius\cite{6,7}. Moreover, the coupling between the center-of-mass motion, of the exciton, and the photon wave vector, that introduces the optical spatial dispersion properties in the polariton model\cite{8}, in principle can give strong radiation-matter coupling also in a single "high quality" quantum well, if the polariton splitting energy is greater than the total broadening of the system. 

When a cluster of N emitters is arranged periodically in an homogeneous dielectric background, with emission wavelength equal two times the spatial periodicity, a new collective coherent state  (super-radiance mode) is present in the system, also called resonant photonic Bragg reflectors (RPBR). Since the  strength of N oscillators concentrate into the super-radiant mode, the radiation-matter interaction increases by increasing the number of  quantum well, and therefore polariton splitting energy can overcome the total broadening of the system (the so called weak/strong coupling transition\cite{4,9,10}).

Recently, the RPBR  becomes an important tool for the control and manipulation of light. In fact, the stopping, storing and releasing light in RPBR was recently proposed by Yang et all.\cite{10}; the former effect is based on the parametric manipulation of photonic band structure, and  practical limitations for its realization, due to the large number of quantum wells present in the cluster ($N > 200$), was also discussed in the same reference.   Moreover, a multilayer system, where exciton energy is close to the transmission peak at high energy side of the stop band was studied experimentally by Askitopoulos et all.\cite{11} as a new tools for tailoring of light-matter interaction, and promising for polariton lasing, which requires no population inversion\cite{12,13}.  
At variance of micro-cavities, where the number of quantum wells are strongly limited, the RPBR, obtained by alternating isotropic materials with different background refraction indices and quantum well resonances, close to the Bragg frequency, require a number of  elementary cells (N=30-60), greater than that present in the BDR microcavity, but no so large as those required  in order to build up the former polaritonic gap\cite{4,5}.

The study of anisotropic photonic crystals, obtained by alternating different transparent uni-axial layers (or isotropic/anisotropic layers), is largely present in the recent literature due to the new interesting optical properties, namely: negative refraction \cite{14,15}, omni-directional transmission\cite{15}  and giant transmission effects\cite{16}.
Recently, in a periodic system obtained by alternating uniaxial bilayers, with different orientation of the in-plane optical  $\hat C$  axis, it was demonstrated that band gap can be opened inside the Brillouin zone, where band curvature indicates the presence of negative refraction\cite{17}. 

The present work is devoted to study the optical response of a new class of resonant hybrid Bragg reflectors (RHBR), where the periodicities, due to the background dielectric constant modulation, and different orientation of the optical $\hat C$  axis on the xy plane are present, and that will be able to combine the optical properties of the isotropic resonant quantum confined systems, namely: super-radiance, strong radiation-matter interaction, high non-linearity, with the interesting optical properties observed in anisotropic multilayers, in particular: high photon density of states at absolute photonic gap, its building up for a rather moderate number of elementary cells, and in general stronger flexibility in the optical tailoring. Finally, optical properties of RHBR will be compared with those shown by analogous RPBR.

Notice, that the recent improvement on plasma-assisted molecular-beam epitaxy\cite{18} have allowed to study multilayer with in-plane optical $\hat C$ axis, and, even if the goal of the total control on $\hat C$ axis orientation in a single multilayer is not very close, the present study of the optical response in RHBR should be a stimulation for studying the optical properties of  systems, interesting for fundamental and applications, as:  i) meta-material topological insulator\cite{19}, ii) repulsive Casimir force\cite{20} and  iii) the so called chiral meso-molecules\cite{21}.   

The aim of the present work is twofold: a) to investigate, for selected physical parameter values, exciton-polariton propagation in a RHBR stacks of  N-Bragg quantum wells under strong radiation-matter interaction for symmetric and asymmetric elementary cell, and, b) to study the dispersion curves of the former systems, in order to investigate the behavior of the so called intermediate exciton-polariton dispersion curves (IDC), in the lowest energy photonic stop band. The latter point is  particularly interesting for light storing as shown in Ref. 10 for isotropic RPBR.
\begin{figure}[t]
\includegraphics[scale=0.55]{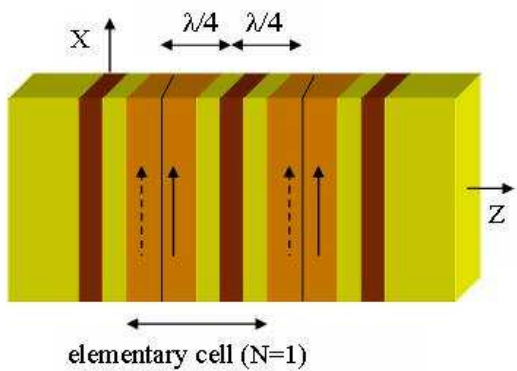}
\caption{(color online) 1D hybrid (isotropic/anisotropic) multilayer. Black  (brown online) layers are the quantum wells, light grey (yellow online) layers are the barriers of the quantum wells and  dark grey (orange online) layers are the uniaxial bilayers. The dashed arrow indicates that $\hat C$ axis could not be parallel to the x axis.}
\end{figure}
The exciton-polariton propagation is computed in a finite RHBR, composed by N elementary cells, where a 2D quantum well is located in an isotropic $\lambda /4$ layer, and the other   layer is composed by uniaxial bilayers, with in-plane $\hat C_\alpha  $  axis ($\alpha  = L,R$), located at left and right side of the uniaxial bilayer, as schematically reported in the Fig. 1 for the case N=1 elementary cell.  Notice, that in all the present calculation we consider a cluster of N elementary cells at $\lambda /2$  for a total of  N+2 quantum wells, due to the insertion of two surface quantum wells (see Fig. 1).

The study of photonic dielectric gap and super-radiant exciton-polariton mode building up, as a function of quantum well number N in a RHBR, is performed by self-consistent calculation and in effective mass approximation\cite{6,7}. Notice, that in the present study the use of non-local self-consistent framework is mandatory for a correct estimation of the radiative self-energy, whose imaginary part, complemented with non-radiative homogeneous broadening (usually derived from the experiments), allows to asses the reaching of the strong coupling regime in the radiation-matter interaction\cite{6} and  gives a quantitative estimation of the absorbance spectra in the so called "high quality" quantum wells, where the non-homogeneous broadening should be negligible.
 In order to obtain an heavy-hole Wannier exciton optically isotropic, in an otherwise anisotropic super-lattice, we  have to minimize the envelope function penetration depths into the uniaxial layers and also the effect of the image potential on the exciton dipole, due to the presence of  different uniaxial layers at both sides of the quantum well (see Fig. 1), as discussed by V. Trongin et all.\cite{22}.   The former requirement is accomplished by considering an heavy-hole Wannier exciton perfectly confined into a 2D quantum well, and clad between two rather large (with respect to the exciton Bohr radius) isotropic barriers, while the latter requirement is obtained by taking the background dielectric constant equal for well and isotropic barriers ($\varepsilon _b  = \varepsilon _w $), and moreover not very different in value from the orthogonal ($\varepsilon _ \bot  $) and parallel ($\varepsilon _\parallel  $) dielectric constants of the uniaxial layers.
\begin{figure}[t]
\includegraphics[scale=0.4]{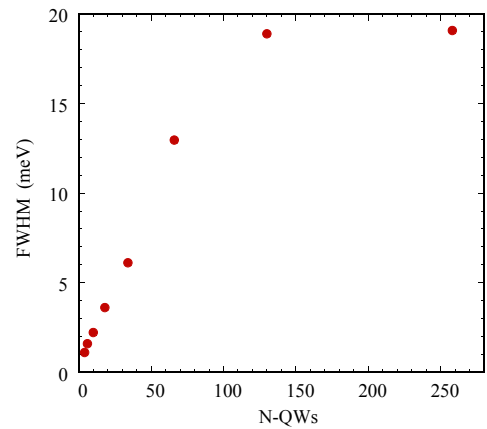}
\caption{ (color online) Full width at half maximum of the reflection spectra of a resonant  Bragg quantum wells at normal incidence in an isotropic bulk as a function of the number N of elementary cells. Parameter values are given in Sec. II of the paper.}
\end{figure}
Now, since the main approximations, embodied in the model calculation, are briefly discussed, let us summarize the plan of the work.
In Sec. II the procedure for computing the optical response of a RHBR, for symmetric and non-symmetric elementary cell, is presented, and also the behavior of the exciton-polariton propagation in RPBR, for homogeneous and periodically modulated dielectric background, are briefly summarized\cite{3,4}. 
In Section III  the optical response for a finite RHBR with in-plane $\hat C$ axis is computed for symmetric and non-symmetric elementary cell. The computation is performed for N=32   elementary cells (and N=34 quantum wells) where strong coupling regime and photonic band gaps are observed for selected values of physical parameters. Their absorbance properties and the behavior of radiative homogeneous broadening and non-radiative  broadening, 
 particular important in optical applications, will be discussed.  In Section IV the dispersion curves of former systems will be computed ($N \to \infty $) for non-radiative broadening value $\Gamma _{NR}  \to 0$. The behavior of the IDC in resonance and out of resonance with Bragg frequency will be discussed at variance of the correspondent isotropic case\cite{4}.
The conclusions will be summarized in the Sec. V.
%
%
\section{THEORY}
  The study of photonic dielectric gap building up and super-radiant exciton-polariton propagation, as a function of quantum well number N, is performed by self-consistent calculation, and in the effective mass approximation along the line of Refs. 4,6 and 7. A standard method based on Green functions and transfer matrix approach is adopted for numerical calculation, and all the formula are explicitly given in the appendix.
Now, let us shortly remind the radiation-matter properties of N Bragg quantum wells in an isotropic background, computed for the same physical parameter values that will be used for the isotropic slab of the hybrid photonic crystal of  Sec. III.

In the present calculation we have used quantum well parameter values, close to those of the InGaAs/AlGaAs/GaAs(001) semiconductor materials, as for istance:   total mass $M = 0.524$, Bohr radius $a_B  = 8.047$nm  and background dielectric constants: $\varepsilon _w  = \varepsilon _b  = 10.24$. In all the cases the used well width is $L_w  = 10nm$. 

  The exciton transition energy  is: $E_{ex} (K_\parallel  ) = E_{ex} (0) + \hbar ^2 K_\parallel ^2 /2M$,  where ${\bf{K}}_\parallel$ is the in-plane center-of-mass wave vector and, in the present calculations, ${\bf{K}}_\parallel   = {\bf{q}}_x $ while $E_{ex} (0) = 1.418eV$ is the variational energy of the exciton confined between infinite potential barriers.     
The homogeneous non-radiative broadening $\Gamma _{NR}$ is chosen close to that is  experimentally observed in high quality quantum wells at rather low temperature: $\Gamma _{NR}  = 0.25 meV$ and $T \approx 10K$. 

Finally, in order to minimizes the Fabry-Perot oscillations, due to the vacuum/semiconductor background dielectric mismatch, that strongly perturb exciton-polariton propagation,  the 1D cluster is clad between two isotropic semi-infinite bulk materials with $\varepsilon _b $ background dielectric constant. Notice that this effect is usually obtained experimentally by an anti-reflection coating deposited on both the surfaces of the sample.
Now, let us briefly summarize the different regimes of exciton-polariton propagation in an 1D cluster of  N quantum wells at $\lambda /2$ separation embedded in a homogeneous bulk  background ($Z \ge 0$) with refraction index $\varepsilon _b  = 10.24$. 

In Fig. 2 the full width at half maximum of the reflectivity is shown as a function of the number N of quantum wells, for resonant wavelength: $\lambda  = 2\pi c/\left( {\omega _{ex} \,\sqrt {\varepsilon _b } } \right) = 2d$, where d is the periodicity and $\hbar \omega _{ex}  = E_{ex} (0)$.  Three different regimes can be observed\cite{4}, namely: the linear behavior, due to the super-radiant mode, at low N quantum well number ($N = 1 \div 80$), the saturation zone, obtained for large quantum well number ($N \ge 130$), where multi-modes polariton propagation are observed (that for $N \to \infty $ will merge in forward and backward polariton Bloch propagation) and an intermediate zone ($80 \le N \le 130$), where the system changes from super-radiance to multi modes behavior.
Notice, that in the zone of super-radiant mode the system can change its behavior from weak to strong radiation-matter coupling when the polaritonic splitting energy, that is a  direct function of  the number N of quantum wells present in the stack, becomes greater than the total (radiative plus non-raditive) homogeneous broadening ($\Gamma _t  = \Gamma _{NR}  + \Gamma _R $). 

\begin{figure*}
\centering
\subfloat[$\alpha = 0^o$]{\includegraphics[scale=0.35]{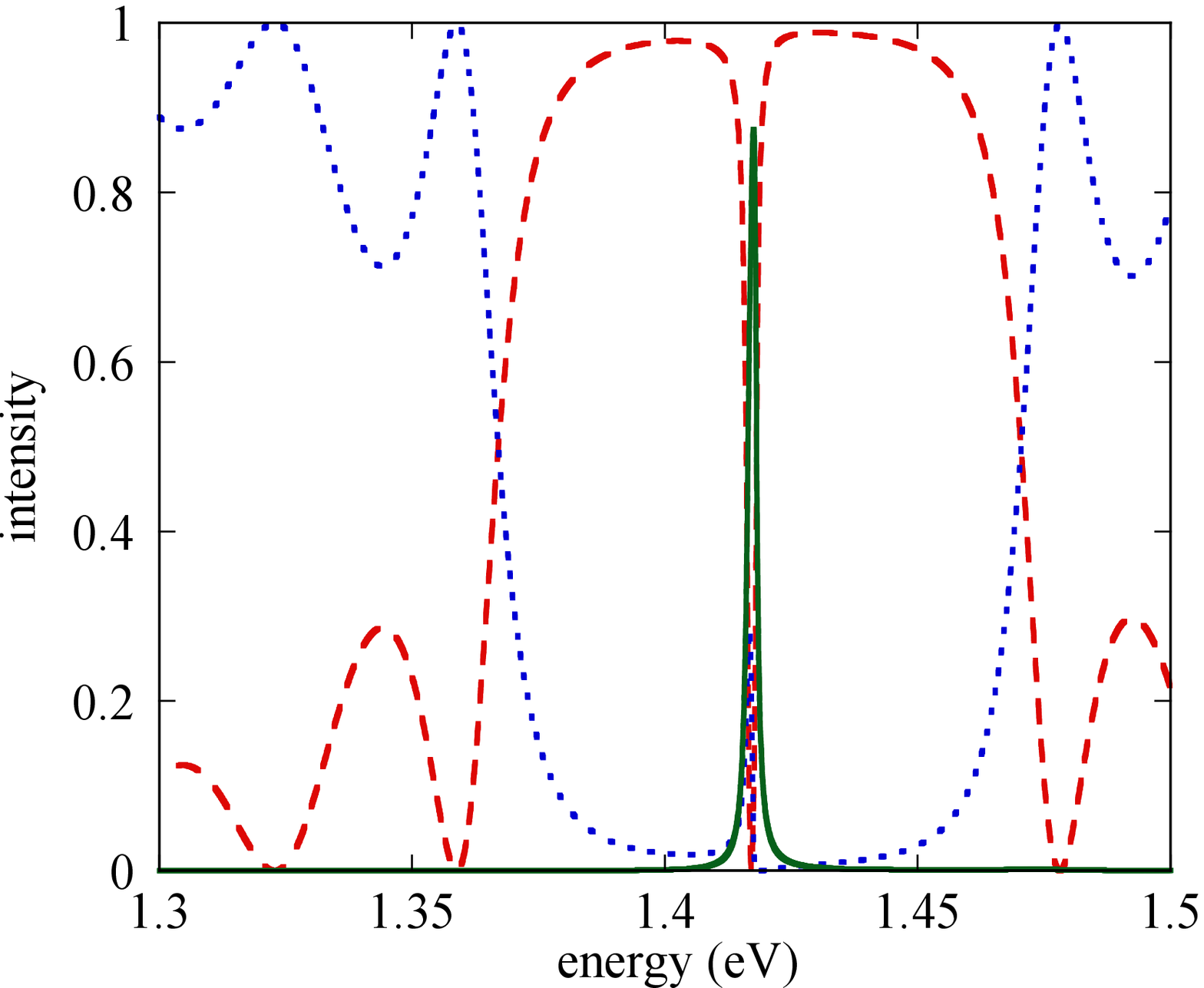}}          
\subfloat[$\alpha = 90^o$]{\includegraphics[scale=0.35]{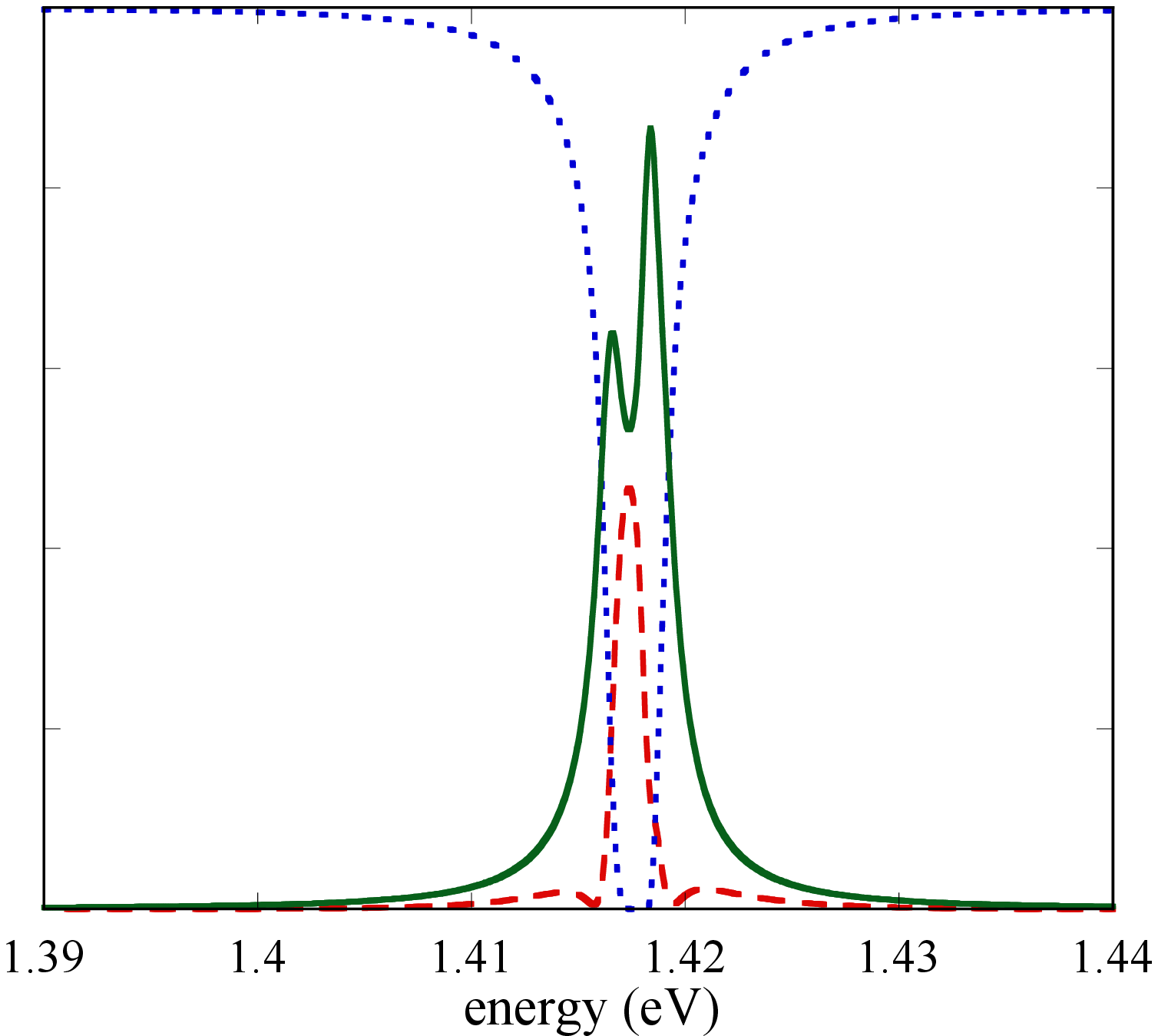}}
\subfloat[$\alpha = 45^o$]{\includegraphics[scale=0.35]{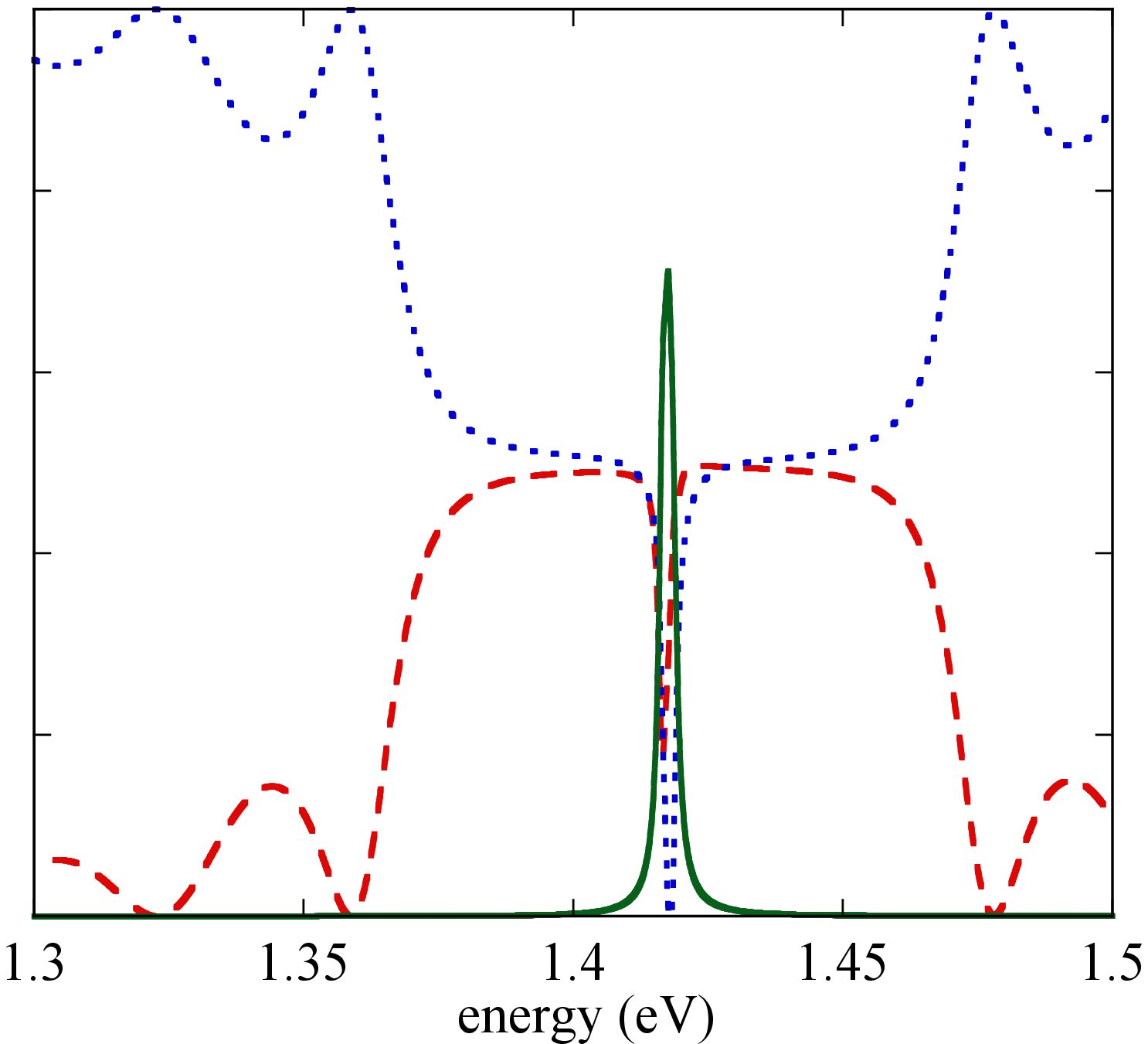}}
\caption{(color online) P-polarized optical response, at normal incidence, of a RHBR (N=32), computed for different $\alpha$ values.  Reflectivity: dashed line (red online); transmission: dotted line (bleu online) and absorbance: solid line (green online).} 
\end{figure*}
At variance of the behavior shown in Fig. 2, where  the polaritonic stop band is due to the Bragg periodicity of the dispersive component of the exciton susceptibility ($N \ge 130$), the photonic stop band due to the isotropic  dielectric background modulation, shows a more fast convergence (for $N \approx 30 \div 60$) and broader stop band in energy also for rather low dielectric contrast $\Delta \varepsilon  = \varepsilon _2  - \varepsilon _1 $.  In fact for $ \varepsilon _2 = \varepsilon _B $ and  $\varepsilon _1  = 8.24$, the stop band width of an isotropic Bragg reflector in resonance with $\hbar \omega _o  = E_{ex} (0)$, computed by the equation: $\Delta \lambda  = \left( {4/\pi } \right)\lambda _o sin^{ - 1} \left( {n_2  - n_1 } \right)/\left( {n_2  + n_1 } \right)$  ($n_i  = \sqrt {\varepsilon _i } $),  is $\Delta \omega _o  \approx 81meV$ , while for homogeneous background we obtain: $\Delta \omega _o  \approx 19meV$. In conclusion, while the exciton-polariton propagation in a resonant Bragg system, with homogeneous dielectric constant background, shows three different well separated regimes as a function of N  (as shown in Fig. 2 for $N = 1 \div 250$), the exciton-polariton propagation in a resonant Bragg system, with background dielectric modulation, show a zone of N-values ($N = 30 \div 60$) where super-radiant and photonic gap regimes are both present in the same cluster.
\section{OPTICAL RESPONSE OF PHOTONIC  HYBRID CLUSTER}
%
%
In the present calculation we consider the optical response in a RHBR, composed by N symmetric elementary cells with z axis along the periodicity and (x,y) in-plane coordinates, for incident wave polarized S or P, and in-plane optical $\hat C$ axis of the uniaxial slab ($\alpha _L  = \alpha _R  = \alpha $) as shown in Fig.  1 for $N = 1$.

As discussed in the introduction, the uniaxial refraction index values are chosen very close to the isotropic refraction index $n_b $, namely: $n_ \bot   = n_b  = 3.2$ and $n_\parallel   = 3.5$ respectively. The former choice emphasizes the role of the orientation of the optical $\hat C$  axis on the xy plane, and makes simpler to compare the total optical response in RHBR with that  observed in equivalent RPBR system.
Notice that the tuning between the exciton energy and the Bragg energy of the photonic stop band in uniaxial periodic multi-layers with symmetric cells can be computed straightforward for $\alpha  = n\,\pi /2$-values ($n = 0, \pm 1, \pm 2,...$), since the dielectric tensor of uniaxial slab becomes diagonal, and therefore, ordinary and extraordinary wave components do not interact. 
For a cluster of N=32 symmetric elementary cells, strong coupling regime is observed\cite{11}, and in order to obtain the Bragg energy $\hbar \omega _o $, of the reflection stop band at
\begin{figure}[t]
\includegraphics[scale=0.42]{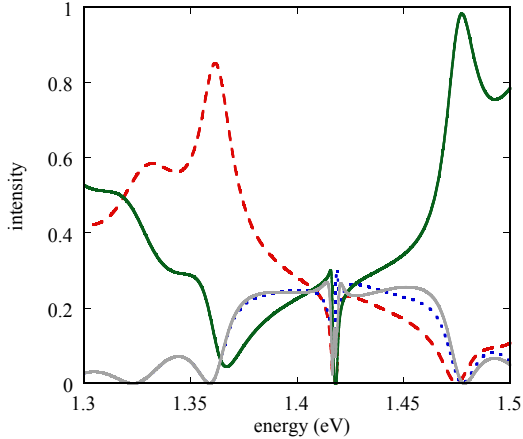}
\caption{(color online) Transmission x-component ($T_{pp}$):  dashed line (red online);  y-component ($T_{ps}$): dotted line (bleu online) and   reflection x-component ($R_{pp}$): dark solid line (green online);  y-component ($R_{ps}$)  light solid line (grey online),  computed for the system of  Fig. 3c.}
\end{figure}
 normal incidence, in resonance with exciton energy ($\hbar \omega _o  \approx E_{ex} (0) = 1.418eV$), for incident P polarized wave we will have to compute the $\lambda /4$ isotropic thickness for wavelength: $\lambda  = 2\pi c/\left( {\omega _{ex} n_b } \right)$, while the total thickness of the$\lambda /4$ uniaxial layer is given for wavelength: $\lambda  = 2\pi c/\left( {\omega _{ex} n_\parallel  } \right)$.
Notice that, due to the optical symmetry of the system\cite{16}, the incident P-polarized optical response, with $\alpha$ orientation of the $\hat C$ axis, shows the same optical spectra than that computed for S polarized incident wave and $\alpha ' = (\pi /2) - \alpha $ orientation.     

In Fig. 3a the total optical response at the normal incidence for P polarized incident wave is shown. We observe a reflection central deep, very  close to the bare exciton energy, in resonance with a rather intense absorbance peak ($I_A  \approx 88\% $), and a rather broad  photonic stop band of $\Delta \omega (q_x  = 0) = 101meV$ (full width at half maximum)\cite{4,16}.  Moreover, since the dielectric modulation of the multilayer is due to the $\varepsilon _\parallel  $ dielectric constant and to the dispersive component of the exciton susceptibility, while for the ordinary wave the multilayer results homogeneous ($\varepsilon _ \bot   = \varepsilon _b $), the band gap disappears in the spectrum computed for $\hat C$ axis orientation along the y axis ($\alpha  = \pi /2$) as is shown in Fig. 3b. In fact, only the normal exciton-polariton optical spectra are present, where the absorbance shows a double peak with two maxima separated in energy of about 1.7meV.
\begin{figure*}
\centering
\subfloat[P-polarization $\alpha _L  = \alpha _R  = 0^o$] {\includegraphics[scale=0.45]{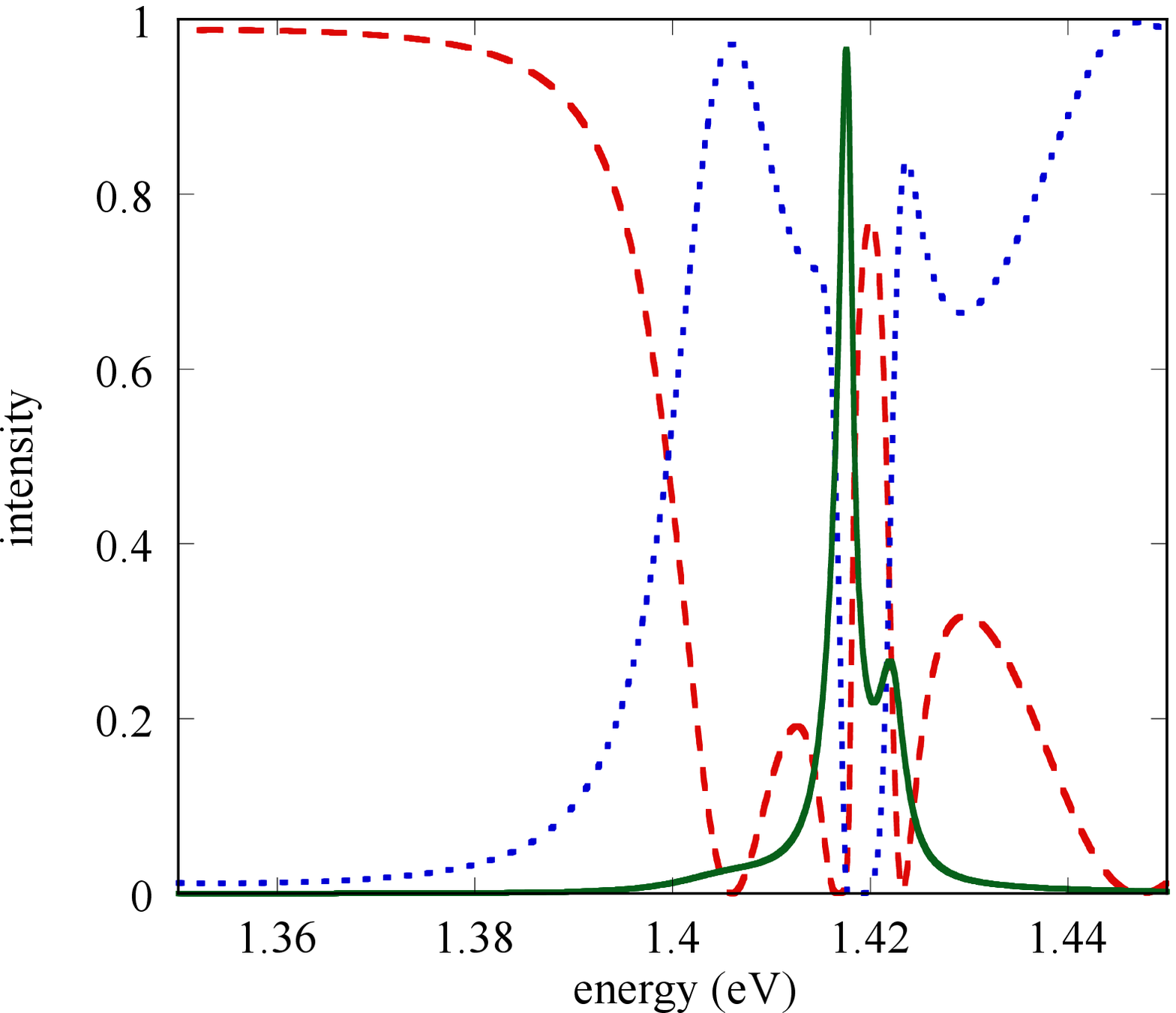}} \quad \quad        
\subfloat[S-polarization $\alpha _L  = \alpha _R  = 0^o$] {\includegraphics[scale=0.45]{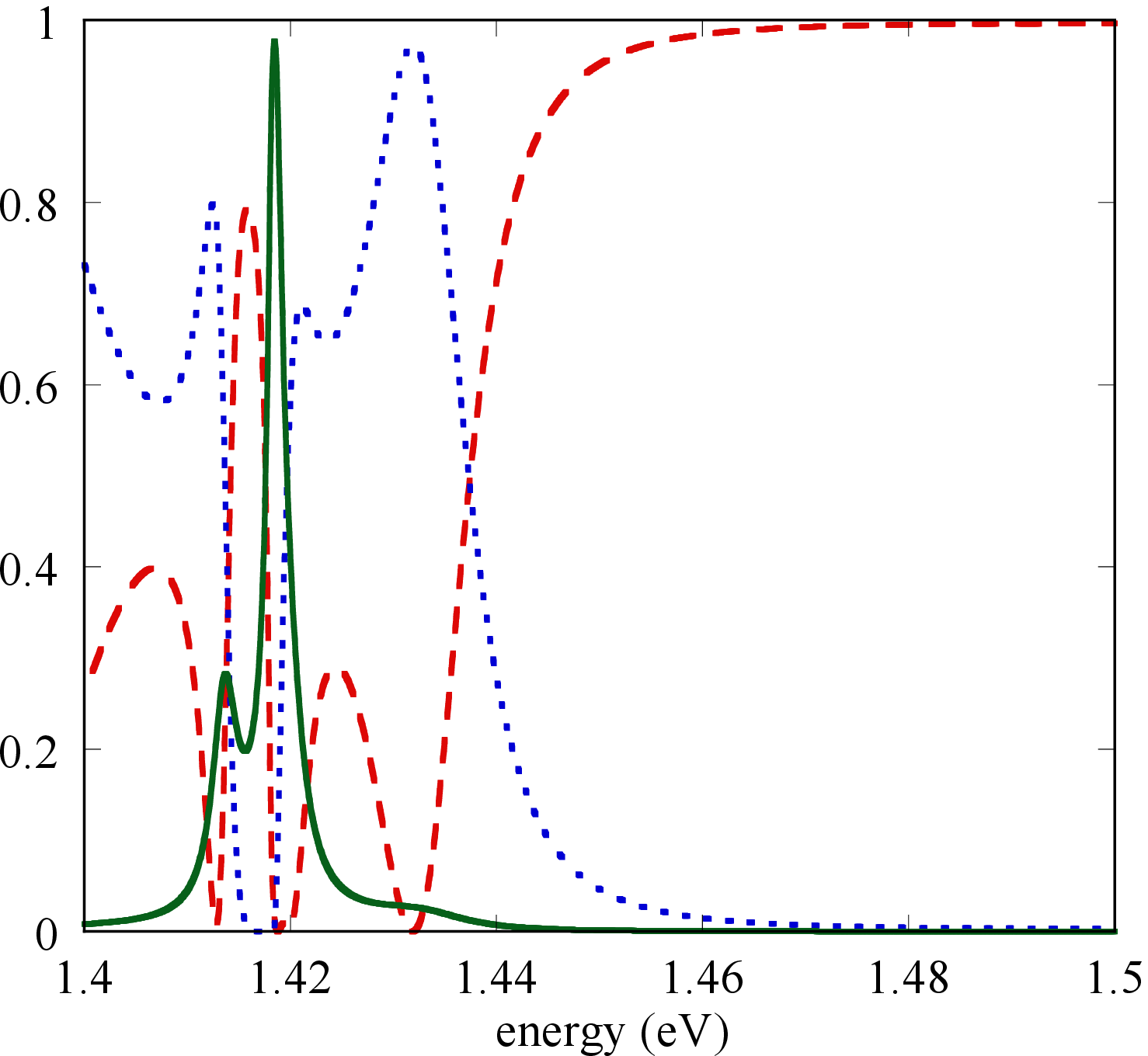}}   
\\
\subfloat[P-polarization $\alpha _L  = 45^o$, $\alpha _R = 0^o$]{\includegraphics[scale=0.45]{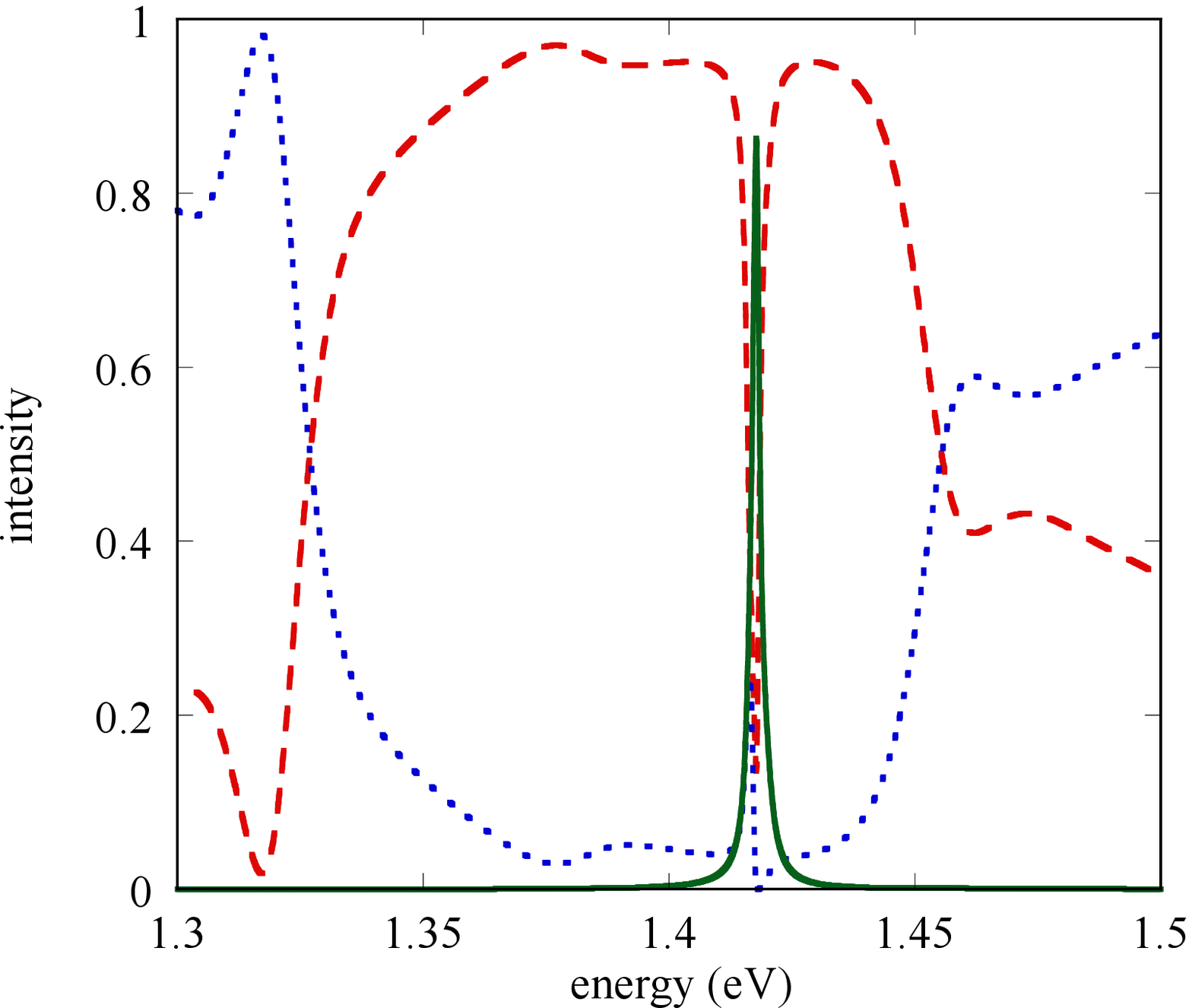}} \quad \quad
\subfloat[S-polarization $\alpha _L  = 45^o$, $\alpha _R = 0^o$]{\includegraphics[scale=0.45]{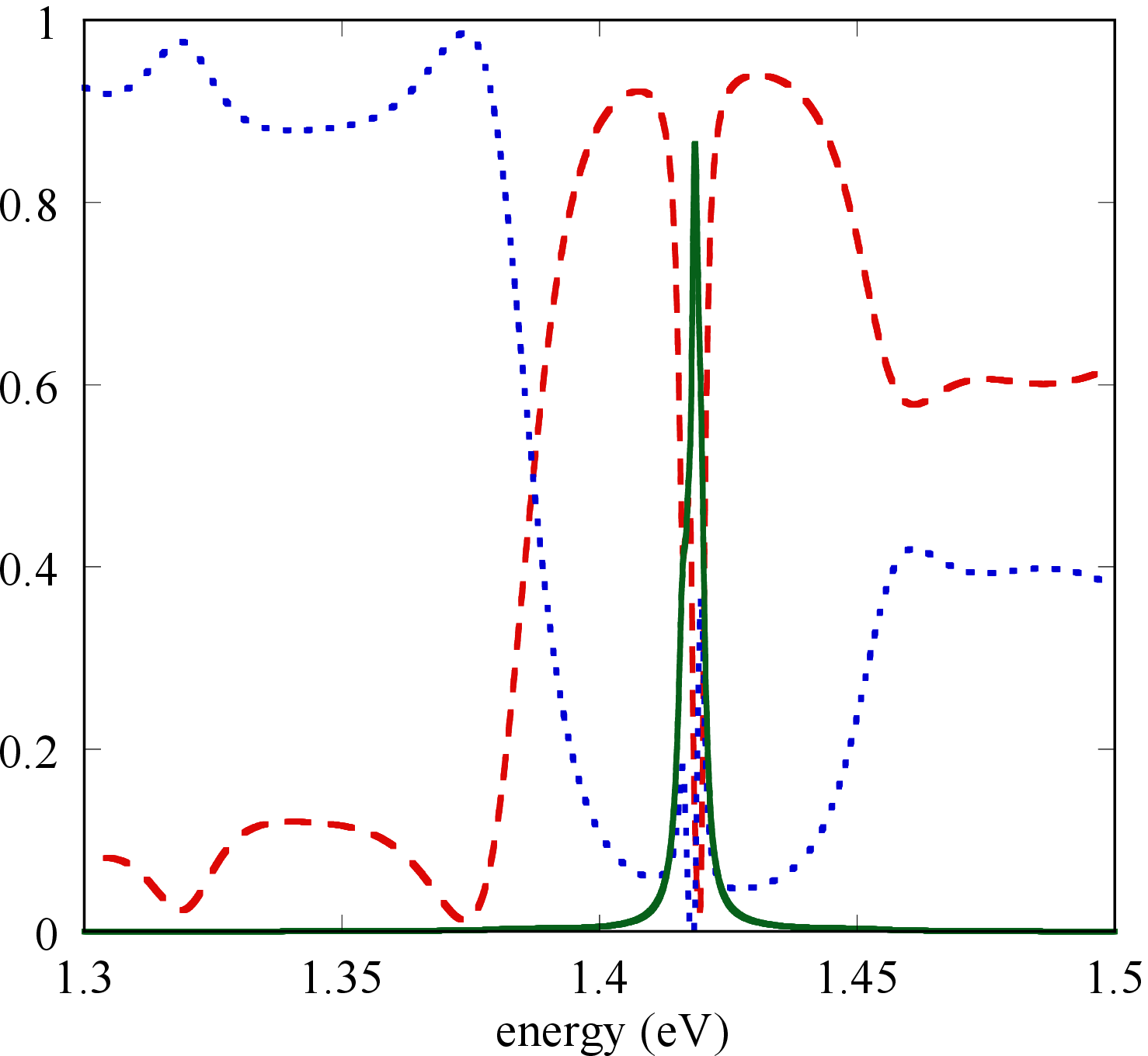}}
\caption{(color online) Linear-polarized optical response, at normal incidence, of a RHBR (N=32), computed for different $\alpha_L$ values.  Reflectivity: dashed line (red online); transmission: dotted line (bleu online) and absorbance: solid line (green online).}
\end{figure*}
Notice, that in the present  kind of RHBR the reflection intensity can be tuned from $0\% $ to $100\% $  by changing the orientation of  the $\hat C$ axis from $\alpha  = 0$ to $\alpha  = \pi /2$, while the line-shape remain unchanged and the  transmission spectrum, close to the border of the reflection stop band, maintains  its high intensity value  ($I_T  \sim 100\% $). 
In fact, the optical response for P incident polarization and $\alpha  = \pi /4$ , where an equal contribution from ordinary and extraordinary waves is expected, is shown in Fig. 3c.  The main optical feature of Fig. 3c is the reflection line-shape that is very similar to the stop band shape except for the intensity that now is close to $50\% $ of reflection, instead of about $100\% $ observed in  Fig. 3a. Analogously, for the transmission line shape in the gap energies, that shows a complementary behavior with respect to the reflection one, with the $50\% $ of the transmission intensity.  Notice that, the former property is, almost verified also by removing the condition $\varepsilon _ \bot   = \varepsilon _b $ (for instance\cite{22}): $\varepsilon _ \bot   = 8.24$.

In order to go a bit deeper in understanding the optical behavior of the former system, let us analyze the optical response as a function of its polarized components. In Fig.  4  for P polarized incident wave the reflection and transmission in-plane components are shown. While the total reflection spectra, for photon energies in the stop band, are obtained by a rather equal contribution of $R_{pp}$ and $R_{ps}$ components, the contribution of the $T_{pp}$ component at total transmission spectra is  essentially localized at low energy side of the stop band, while the $T_{ps}$ contribution is at high energy side. Obviously, for S polarized incident wave the pp and sp transmission  components must be exchanged.

In conclusion, in symmetric RHBR, with $N = 32$, the absorbance spectrum clearly show two components due to IDC bands, and moreover, for selected values of the physical parameters, it is possible to determine the orientation of  the in-plane $\hat C$ axis  by reflection and/or transmission intensity measurements, and also the polarization of the incident wave by analyzing the polarization of  the transmission peaks at low and high energy side of  the photonic stop band. Usually, the optical tailoring of RHBR require the solution of the so called inverse problem, while in the present case these informations can be deduced directly from a quantitative analysis of the optical response.
Now, let us consider a bit more complex system than before obtained by substituting the uniaxial layer with an uniaxial bilayer of equal thickness, with the optical $\hat C_\beta  $ axis (with $\beta  = L,R$) oriented along x-axis for the right layer ($\alpha _R  = 0$) and in-plane for the left layer ($0 \le \alpha _L  \le \pi /2$) respectively. Notice, that in this case the elementary cell is asymmetrical, except in the limit of $\alpha _L  \to 0$, and therefore we will have to distinguish between forward and backward optical response.
 
In the present computation a different choice than  before is adopted for $\varepsilon _ \bot  $ in order to accomplish the following two conditions: $\varepsilon _ \bot   < \varepsilon _b  = \varepsilon _w  < \varepsilon _\parallel  $ where $\varepsilon _ \bot   = 8.24$ and $(\varepsilon _\parallel   + \varepsilon _ \bot  )/2 \approx \varepsilon _b $, moreover an equal $\lambda /4$ thickness value, computed for background dielectric constant $\varepsilon _b $, is chosen for isotropic and anisotropic layers, in resonance with bare exciton energy ($\hbar \omega _o (0) \approx E_{ex} (0)$).
\begin{figure*}
\centering
\subfloat[P-polarization $\alpha _L  = 90^o$  $\alpha _R  = 0^o$] {\includegraphics[scale=0.45]{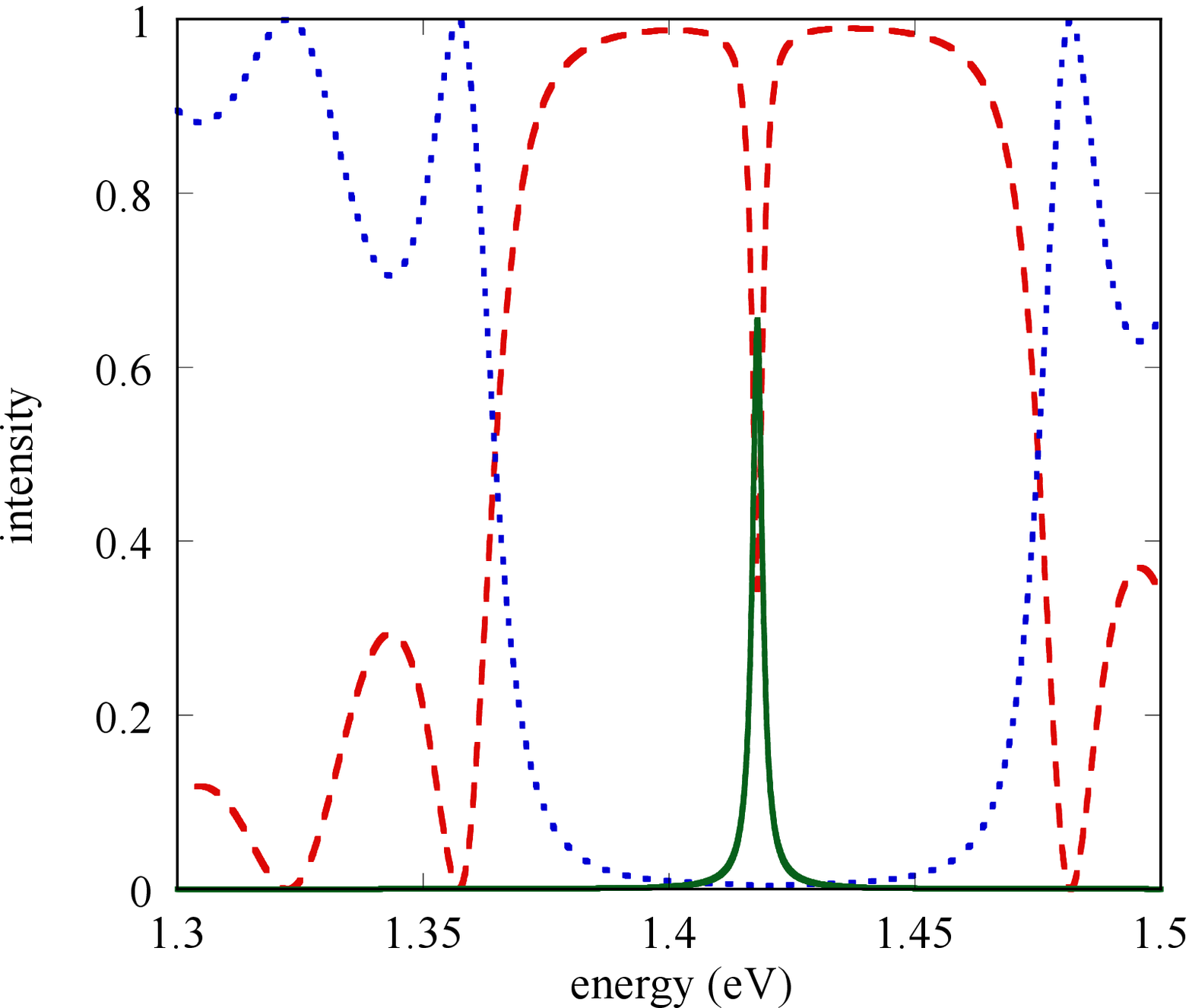}} \quad \quad           
\subfloat[S-polarization $\alpha _L  = 90^o$  $\alpha _R  = 0^o$] {\includegraphics[scale=0.45]{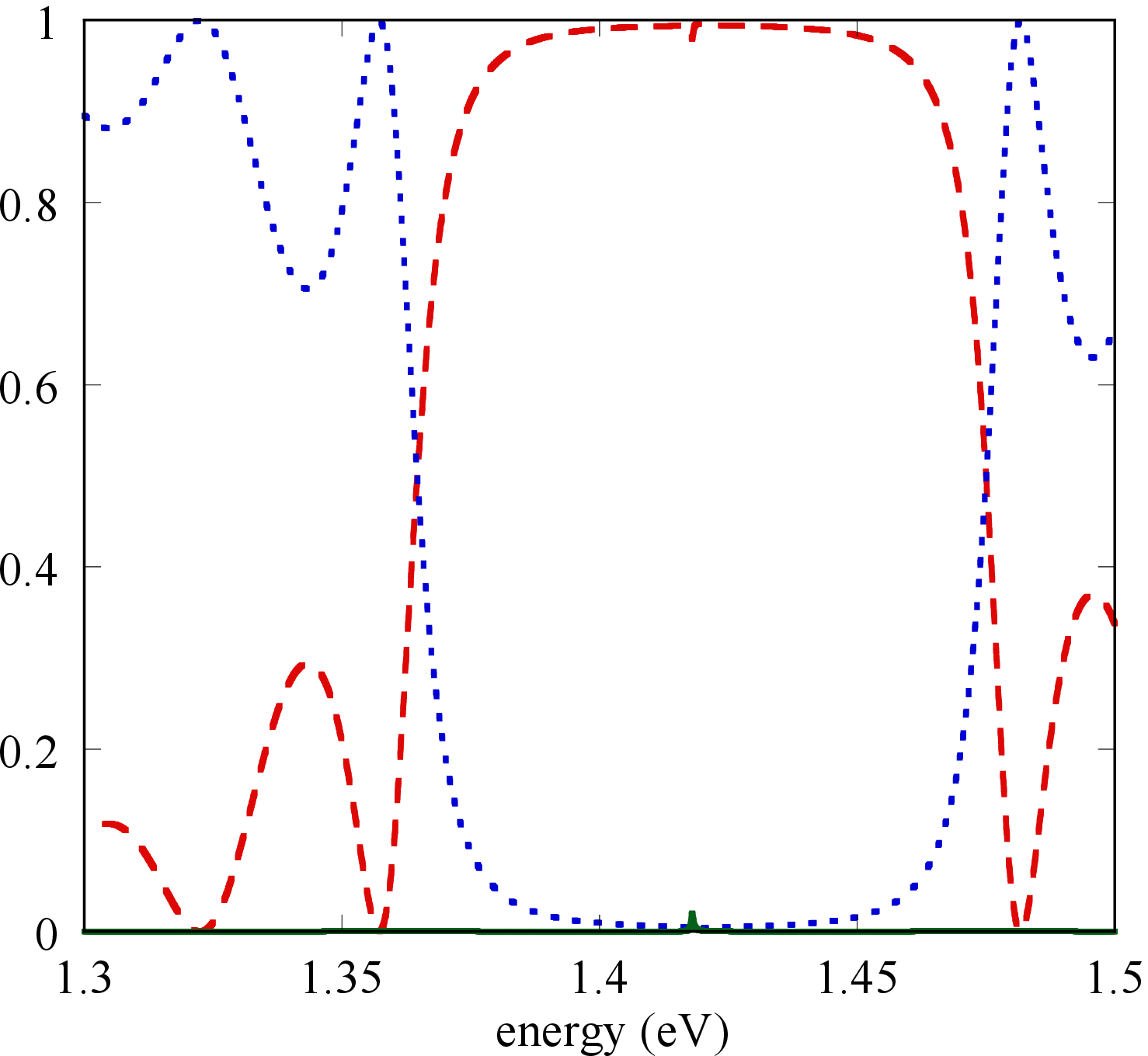}}   
\caption{(color online) Optical response, at normal incidence, of a RHBR (N=32).  Reflectivity: dashed line (red online); transmission: dotted line (bleu online) and absorbance: solid line (green online).} 
\end{figure*}
First of all, let us to compute total reflectivity and adsorption for S and P incident waves  for a multilayer with N=32 elementary cells in the limit of symmetric elementary cell ($\alpha _L  = \alpha _R  \to 0$). In this case the system is similar to that discussed before, except for the thickness of the uniaxial bilayers that now is in resonance neither  with $\lambda /4$ computed by $n_\parallel  $ nor with that computed by $n_ \bot  $ values.  The numerical results are shown in Fig. 5a and 5b for P and S incident polarization respectively, where two well shaped reflection stop bands can be observed at opposite energy sides of the exciton absorption peak, and moreover in the absorption spectra a double peak is present.   
Now, by increasing the $\alpha _L $ value (with $\alpha _R $ taken constant and equal zero) the forward optical response shows that the two stop bands move in the opposite directions in energy,  towards the exciton energy value, that will be reached for  $\alpha _L  = \pi /4$ as shown in Fig. 5c  and 5d for P and S incident wave polarization respectively.  Notice that for the former $\alpha _L $ value a complete mixing between ordinary and extraordinary waves are expected, and a well formed photonic stop band ($\Delta \omega (0) \approx 57meV$) in resonance with bare exciton energy are present for both the polarizations. 
Notice, that at first sight the reflection spectrum of Fig. 5c could be interpreted as a very broad reflection band $\left( {\Delta \omega (0) \approx 114mev} \right)$ where the Bragg energy seems not in resonance with the exciton energy, but that this interpretation is not correct can to be demonstrated from the location in energy of the two side bands\cite{4} present in the exciton absorbance spectrum (not reported here) and from the dispersion curves that will be discussed in the next section.

\begin{figure}[t]
\includegraphics[scale=0.45]{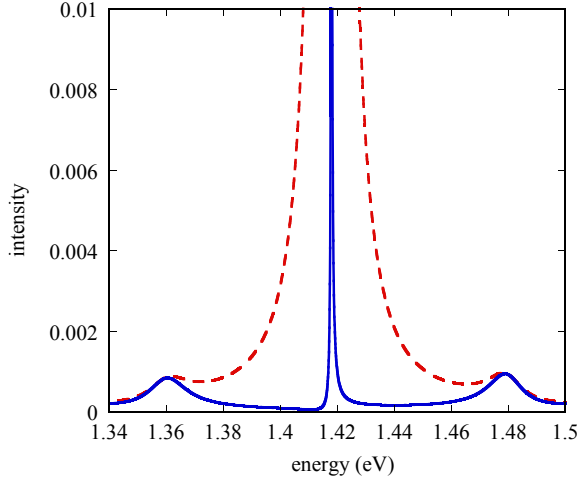}
\caption{(color online) Absorbance spectra of the system of Fig. 6a (dashed line) and Fig. 6b (solid line) in enlarged scale of intensity ($A = 0 \div 0.01$).}
\end{figure}
Finally, for  $\alpha _L  = \pi /2$ two photonic band gaps, with Bragg energies in resonance with the exciton energy $\left( {\Delta \omega (0) \approx 103mev} \right)$,   are shown in Fig. 6a and 6b for both the polarizations. Notice, that in Figs. 6a and 6b the two spectra show exactly the same line shapes, due to the $\alpha _L  = \pi /2$ condition\cite{17}, except for photon energies very close to the exciton transition energy, where exciton absorptions show rather different intensities, but analogous line shapes, for S and P incident wave
 polarizations as shown in  Fig. 7. Since the forward absorption spectra show the same rather small absorbance intensity for S and P incident wave polarizations under the condition $\Gamma _{NR}  \to 0$  (not reported here), we can conclude that in the present case the homogeneous non-radiative broadening affects essentially P polarized incident wave spectrum, while gives a very small contribution to S-polarized one.
\begin{figure}[t]
\includegraphics[scale=0.45]{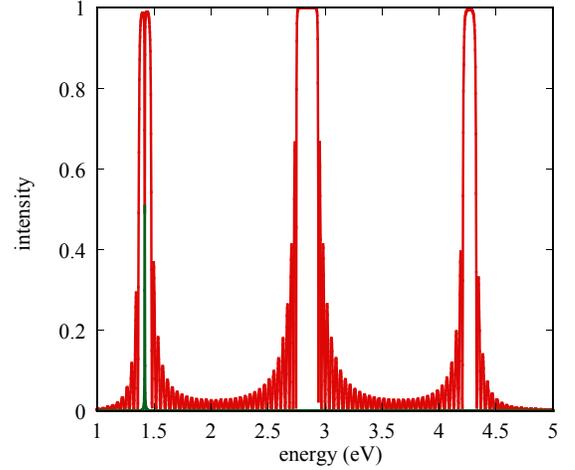}
\caption{(color online) The same optical response of Figs. 6a shown in a wider energy scale.}
\end{figure}
The former effect is due to the asymmetry of the elementary cell, as it could be easily verified by observing that S and P absorption spectra exchange their intensities, by exchanging forward with backward in the exciton-polariton propagation. Moreover, the former interpretation is in complete agreement also with qualitative considerations derived from the computation performed in an equivalent isotropic RPBR.

In conclusion, a quantitative estimation of non-radiative homogeneous broadening can be obtained from experimental optical response performed in RHBR by adopting an asymmetric elementary cell. Notice that, the former interesting result is easily generalized to the RPBR with asymmetric elementary cell.

Now, before to discuss the dispersion curves ($N \to \infty $) of the former system, for sake of completeness, let us show the forward optical response of  P incident wave polarization of Fig. 6a, reported in a much larger range of photon energies where also higher energy stop bands are present (see Fig. 8). In this
 case the absorbance peak, correspondent to the IDC, at the Bragg energy of the lowest energy stop band, is also clearly shown.  For S incident wave polarization we should observe the same line shape as for P polarization, except for intermediate exciton-polariton absorbance peak that in this scale of intensities should be hardly observed.
%
%
\section{EXCITON-POLARITON DISPERSION CURVES}
In order to go a bit deeper in understanding the exciton-polariton propagation in the former hybrid periodic system with asymmetric elementary cells the computed dispersion curves (for $N \to \infty $ and $\Gamma _{NR}  \to 0$) are shown in Fig. 9a for  $\alpha _L  = \pi /4$.  Notice, that the optical response for a cluster of  N=32 asymmetric elementary cells are reported in Fig. 5c and 5d for incident waves polarized P and S respectively.
\begin{figure*}
\centering
{\includegraphics[scale=0.55]{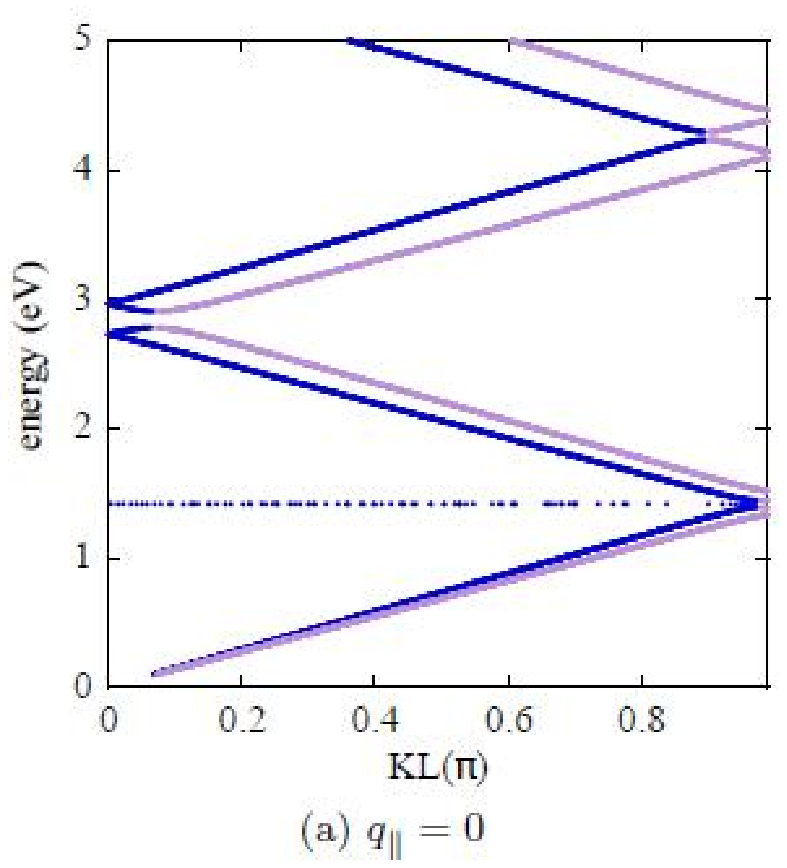}}          
{\includegraphics[scale=0.55]{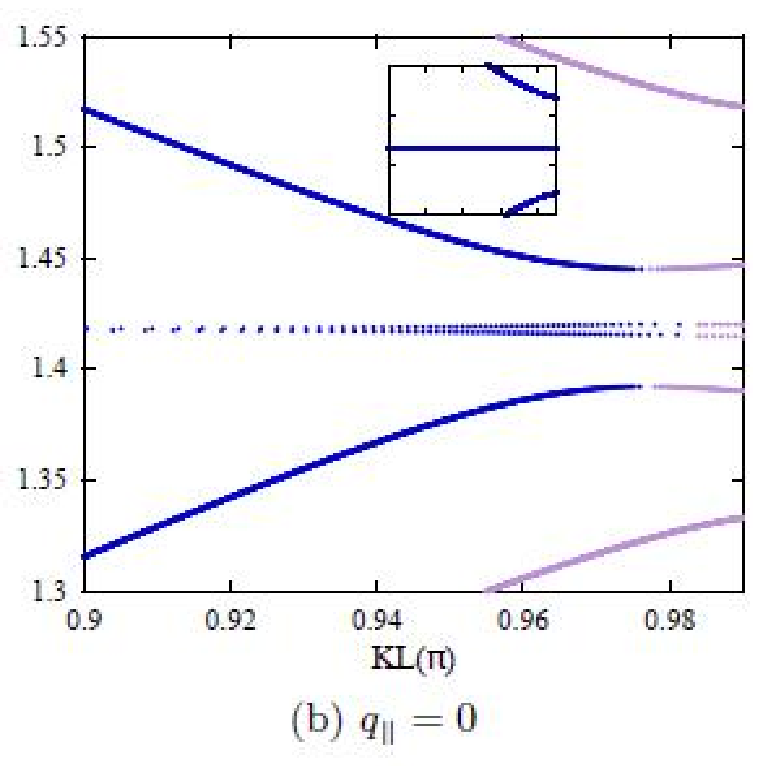}}
{\includegraphics[scale=0.55]{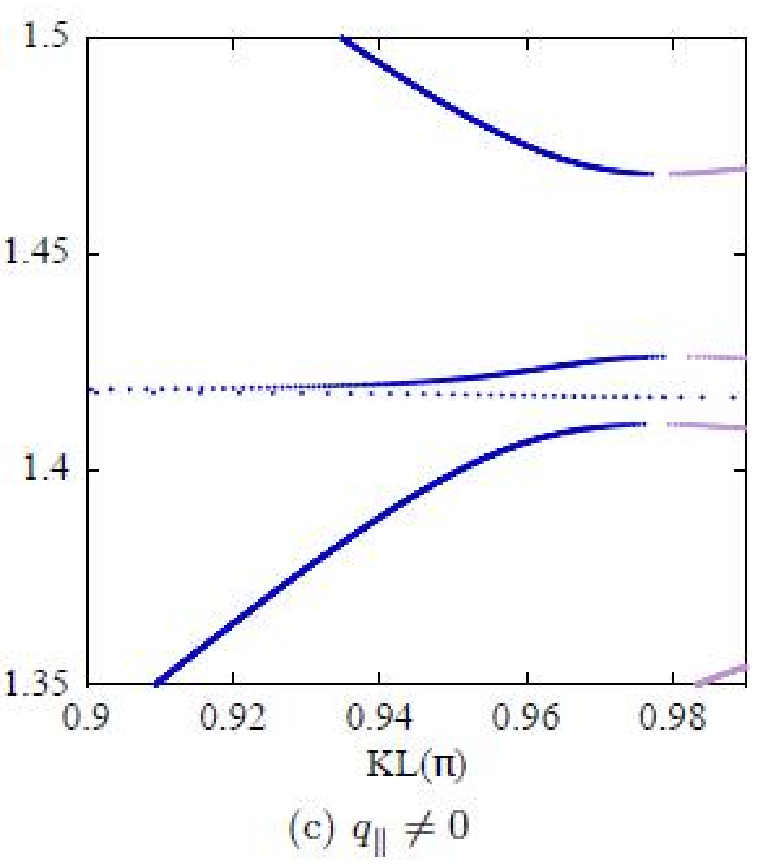}}
\caption{(color online) Dispersion curves  of a periodic system with the same parameter values of the former Fig. 3C, computed in the limits: $N \to \infty $ and $\Gamma _{NR}  \to 0$.  Figs. b and c are shown in a reduced energy range around the  Wannier exciton energy. x-polarized eigenvalues are represented by  light  (mauve online) dots and y-polarized by dark (bleu online) dots.} 
\end{figure*}
The photonic energy gaps drop inside the Brillouin zone, as clearly observed in the second and third gap of the picture, and this is a rather general property of RHBR with in-plane  -axis, due to the anisotropic properties of the elementary cell as discussed in Ref. 17.   
\begin{figure*}
\centering
{\includegraphics[scale=0.65]{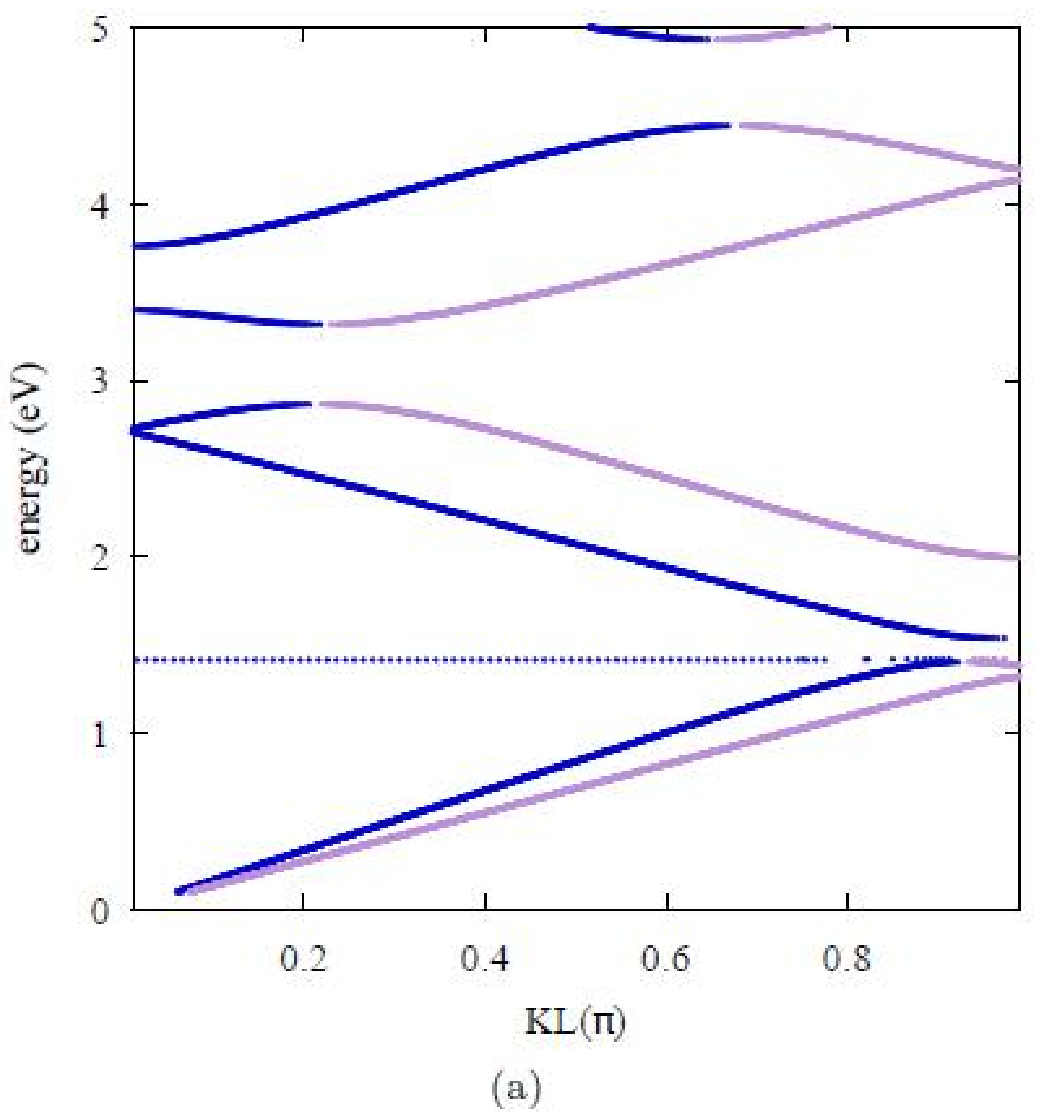}} \quad \quad          
{\includegraphics[scale=0.65]{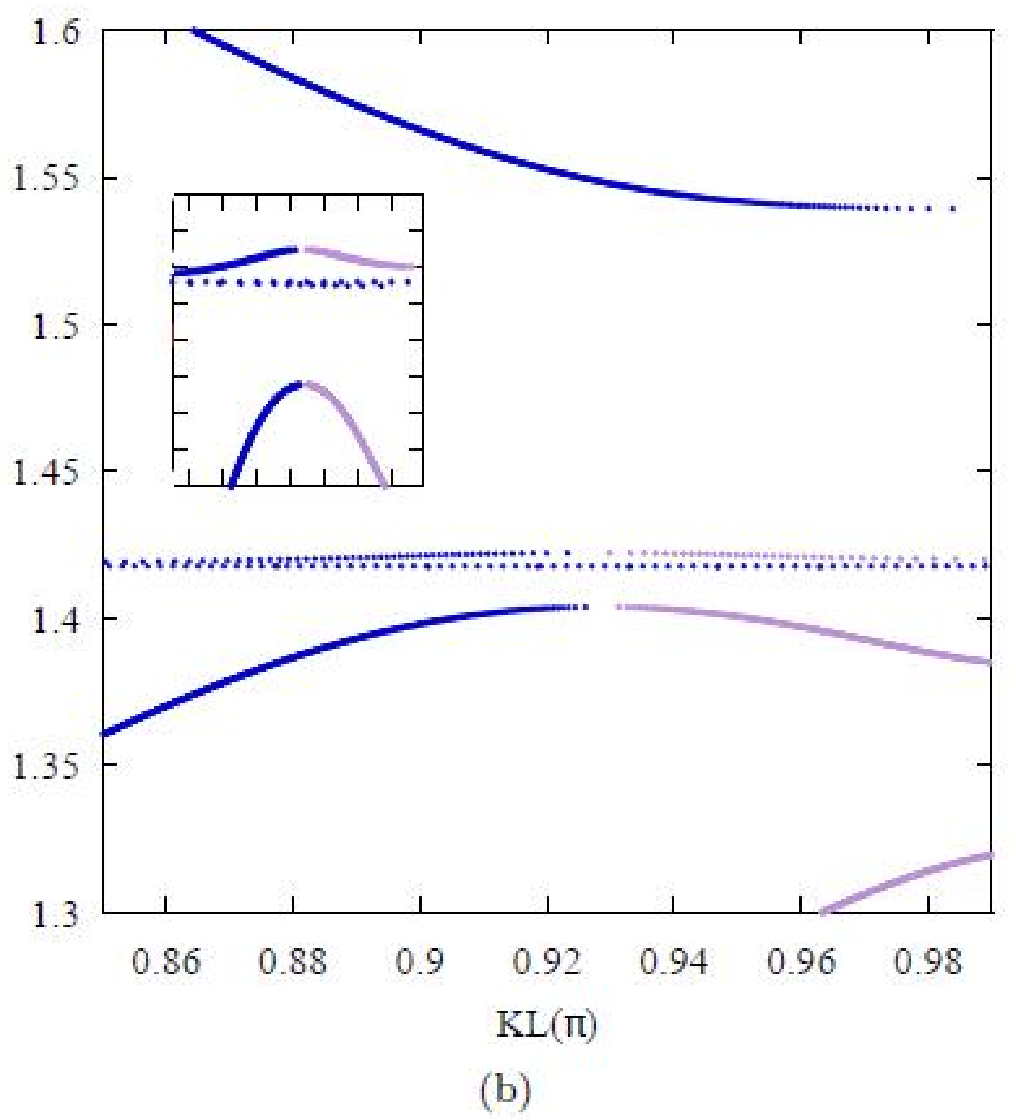}}
\caption{(color online) Dispersion curves  of a periodic system, with the same cell geometry of the former  Fig. 9a, but computed by using $\varepsilon _\parallel   = 13$  and  $\varepsilon _ \bot   = 2.25$.  Fig. b is shown in a reduced energy range around the  Wannier exciton energy. x-polarized eigenvalues are represented by  light  (mauve online) dots and y-polarized by dark (bleu online) dots.} 
\end{figure*}
In the lowest energy gap of Fig. 9a an IDC, close to the unperturbed exciton energy, is also reported. This feature, that is present also in isotropic Bragg quantum wells\cite{4}, characterizes the exciton-polariton propagation and is promising for optical applications; in fact, its rather constant behavior in energy supports zero polariton group velocity, therefore a trapped light in the RHBR could be obtained if this property will be maintained in whole the  Brillouin zone as discussed, in Ref. 10, for isotropic RPBR.  

In Fig. 9b the first gap is shown in an enlarged energy scale. Notice, that the first photonic gap $\left( {\Delta \omega (0) \approx 57meV} \right)$,  is in very good agreement with that observed in the cluster of Fig. 5d, for  S incident wave polarization (dark curves), while for P polarization (light curves) a further high reflection zone, at lower photon energies, is present. Notice, that the lowest energy dispersion curve interacts, with the exciton-polariton IDC, giving rise to the asymmetric behavior of the optical response of Fig. 5c, as discussed in the former section.
 
Very close to the boundary of the Brillouin zone, the lower and upper band of the absolute photonic gap show a strongly deformed behavior, with respect to the normal parabolic one, that should give high polariton density of states, and this property is related with the so called "gigantic transmission band edge resonances\cite{16}".  Moreover, two IDC are present in the gap, that are very close to the unperturbed exciton energy, except for quasi-momentum K-values, in corrispondence of the direct gap, where the interaction with lower and upper photonic dispersion curves removes the degeneracy and therefore they show an energy separation as large as $\Delta \omega  \approx 5.5meV$. The former energy splitting between the two IDC are due to the repulsion between the high density of states of the upper photonic dispersion curve of the gap with the lower energy IDC that show the same symmetry, and between the lower dispersion curve with the upper energy IDC\cite{11}. 
Notice, that for $\alpha  = \pi /2$, where band structures become degenerate for P and S wave polarizations\cite{17}, these two curves recombine in one IDC two fold curve (see the inset of Fig. 9b), that is rather constant in energy in the whole  Brillouin zone, since the interaction between upper and lower photonic branches and the two IDC becomes strongly reduced due to the increasing of the energy gap ($\Delta \omega (0)$ changes from  about 57meV to about 90meV). Moreover, the band gap moves to the border of  Brillouin zone and shows a normal quadratic behavior .  Therefore, the possibility of opening or closing the window of $\Delta \omega  \approx 5.5meV$ by different orientation of the left $\hat C$ axis is very promising for studying the light storing effect in RHBR. In fact, at variance of an analogous isotropic case discussed in the Ref. 10, where the switch on/off is obtained by dynamical Stark effect, in the present system  the switch is obtained by a linear renormalization of the photonic gap, and this is the most interesting result of the present section.  Moreover, it is well known that dynamical Stark effect decreases the exciton oscillator strength, and therefore the system can undergo an unwanted  transition from strong to weak coupling regime.

  In principle, to remove the degeneracy between the two IDC on the energy gap should be obtained also for non-normal incidence propagation in the RHBR. In fact, let us consider the dispersion curves for non-zero in-plane wave vector $\left( {q_x  = q_\parallel   \ne 0} \right)$ computed for incident angle $\theta  = 10^o$. In this case, while the energy shift of the exciton center-of-mass is negligible small  $\left( {\hbar ^2 q_x^2 /2M \approx 6.7\mu eV} \right)$, the shift, due to the longer path of the light in the multilayer, is clearly observed for $\alpha _L  = \pi /4$  (see fig 9c).  
In fact, the direct polaritonic gap changes from $\Delta \omega (0) \approx 52meV$ to $\Delta \omega (q_x ) \approx 42meV$, and  the top of the lower photonic band shift in energy of: $\hbar \left[ {\omega _v (q_x ) - \omega _v (0)} \right] = 33.3\,meV$. 
Notice, that in this case only three peaks are present in the optical spectra if the relationship $\Delta \omega (q_x ) \gg \Delta \omega _p $ is verified; moreover, while the energy splitting between the upper IDC and lower polariton branch interaction is about $\Delta \omega _p  \approx 10\,meV$  the lower IDC seems rather unperturbed, since the upper polariton branch is rather higher in energy. Notice, that the former result is in rather good agreement with analogous result obtained in isotropic RPBR in the Ref.  11 where the interaction is between upper photonic branch and the lower IDC.

Finally, we have computed the dispersion curves for the same structural system discussed before, but with stronger dielectric contrast in the uniaxial bilayer ($\varepsilon _\parallel   = 13$ and $\varepsilon _ \bot   = 2.25$). In Fig. 10a  more pronounced energy gaps are present inside the Brillouin zone, therefore, for exciton energy close to one of these photonic gaps an anomalous exciton-polariton propagation could be present\cite{16,17}; moreover, also in this case, due to the shift of the photonic band gap, two IDC, close to the top of the first optical valence band, are observed. 
In Fig. 10b the first energy gap is shown in an enlarged energy scale and with this resolution an indirect absolute photonic gap is observed ($\Delta \omega (q_x  = 0) \approx 119.2meV$). In a further enlarged energy scale (see the inset of Fig. 10b), the former pattern looks rather symmetric with respect to the maximum of lower photonic band energy. Two dispersion curves are present in the photonic gap with strong different behaviors, namely: one dispersion curve is a rather unperturbed and shows a negligible group velocity till very close to the boundary of the Brillouin zone\cite{4}, while the second curve shows strong distortion due to the repulsion with the top of the lower photonic band,  its group velocity shows different sign with respect to the maximum of the dispersion curve\cite{17}. Since also in the present case $\Delta \omega (q_x  = 0) \gg \Delta \omega _p $, a pattern of three polariton dispersion curves is confirmed in the interaction zone\cite{11}.

In conclusions, hybrid isotropic/anisotropic resonant photonic crystals seems a rather promising class of meta-materials, with respect to the isotropic multilayers, because strongly enlarges the possibilities of the optical  tailoring for fundamental studies and device applications\cite{21}.
\section{CONCLUSION}
In the present work, the non-local optical response of a RHBR cluster with in-plane $
\hat C$ axis, composed by N symmetric and asymmetric elementary cells, is computed by self-consistent calculations and in the effective mass approximation. The non radiative homogeneous broadening value, that in the model calculation is the only parameter derived from the experiments, is taken coherently with those adopted for the so called "high quality" quantum wells.

The optical response of a RHBR composed by   symmetric elementary cells shows strong radiation-matter coupling and, for selected values of the physical parameters, allows to extract, from reflection and/or transmission intensities, the orientation of the in-plane $\hat C$ axis, and from polarized transmission spectra the polarization of the incident wave. Notice that all these information can also be derived from the solution of the inverse problem. Moreover, the problem of tuning exciton transition energy and Bragg energy of the stop band in a cluster of hybrid multi-layers is also addressed. 

The RHBR with asymmetric elementary cell are obtained by substituting the uniaxial layer of the former system with uniaxial bilayer with different orientation of the optical   axis (see Fig. 1). We show theoretically that, for special values of in-plane $\hat C$ axis orientations, the absorption spectra of this system can give a direct quantitative estimation of non-radiative homogeneous broadening of the exciton-polariton in "high quality" quantum wells. Notice, that the former parameter, as underlined before, strongly affects the results of  the self-consistent calculation of the optical response in RHBR.

Finally, the computation of the dispersion curves in RHBR, with asymmetric elementary cell, has allowed to investigate the exciton-polariton propagation and/or localization, due to the presence of two IDC in the lowest energy gap, very close in energy to the unperturbed exciton dispersion curve, that show a strong different behavior as a function of  $\hat C$  axis orientation. We guess that the former property is promising for optical device realizations.
\appendix
\section{EXCITON-POLARITON MODEL IN A 2D QUANTUM WELL}
The study of exciton polariton propagation, in semiconductor materials, requires the solution of the Maxwell equation:         
\begin{equation}
\boldsymbol{\nabla }  \times \boldsymbol{\nabla }  \times {\bf{E}}(R;\omega ) = q^2   \varepsilon _b {\bf{E}}(R;\omega ) + 4\pi {\bf{P}}_{ex} (R;\omega )
\end{equation}
where $q = \omega /c$,   $ {\bf{P}}_{ex} (R;\omega ) $ is the nonlocal linear polarization vector: 
\begin{equation}
{\bf{P}}_{ex} (R;\omega ) = q^2 S_{ex} \Psi (R)\int\limits_{ - L{}_w/2}^{L_w /2} {dR'} \,\Psi^ *  (R'){\bf{E}}(R';\omega )
\end{equation}
and $\Psi $  is the exciton envelope function.

  In isotropic material we can choose Cartesian coordinates with the Z axis perpendicular to the surface plane (X,Y) and, for the cylindrical symmetry of the quantum well, we can adopt mixed coordinates $(K_\parallel  ,R)$
 for the motion of the exciton centre-of-mass.  In this case the exciton function  can be written as:
\begin{equation}
\Psi (r,R) = N \varphi (r)e^{iK_\parallel  R_\parallel  } 
\end{equation}
where $r = \sqrt {\rho ^2  + \left| {z_e  - z_h } \right|^2 } $. 
For a 2D heavy hole Wannier exciton, perfectly confined between infinite potential barriers,  $\varphi (r)$ ca be written as a two subbands variational envelope function \cite{23}:
\begin{equation}
\varphi (r) = N_{ex} \cos \left( {\frac{{\pi \,z_e }}{{L_w }}} \right)\cos \left( {\frac{{\pi \,z_h }}{{L_w }}} \right)\,e^{ - \rho /a_{ex} } 
\end{equation}

As it is well known, the solutions $E_\beta$ $\left( {\beta  = x,y} \right)$  of Eq. (A1) can be obtained by the solutions $E_\beta ^o$  of the corresponding   homogeneous part (${\bf{P}}_{ex}=0$)  combined with a solution of the heterogeneous equation (see, for instance,  Refs. 4 and 6), namely:  
\begin{eqnarray}
E_\beta  (\omega ,Z) = E{}_\beta ^o (\omega ,Z) -
\tilde q{}_\beta ^2 S_{ex} (\omega ,q_x ) \nonumber \\
\quad \quad \times \int\limits_{ - L{}_w/2}^{L_w /2} {dZ''G_w^{\left( \beta  \right)})} \,\Psi^ *  (Z'') \nonumber \\
\times \int\limits_{ - L_w /2}^{L_w /2} {dZ'\;\Psi (Z')} \,E_\beta  (\omega ,Z')
\end{eqnarray}
where $G_w^{\left( \beta  \right)}$  is the  Green function. Notice that, we have embodied all the different normalization constants \cite{24}, of the  Green function, into the coefficient $\tilde q_\beta ^2 $ that assumes the values: $\tilde q_x^2  = k_w^2 /\varepsilon _w $ and $\tilde q_y^2  = \omega ^2 /c^2 $ respectively. Therefore, the Green functions become the same for  both x and y components:
\begin{equation}
G_w (z,z') = \frac{{e^{iq_w \left| {z - z'} \right|} }}{{i2q_w }}
\end{equation}

All the energy dependence is contained in the quantity:
\begin{equation}
S_{ex} (\omega ) = \frac{{S_o (\omega )}}{{E^2 (K_\parallel  ) - \hbar ^2 \omega ^2  - i2\hbar \omega \Gamma _{NR} }}                
\end{equation}
 where $S_o (\omega ) = 4\pi g\hbar E_K e^2 /\omega m_o $ and $E\left( {K_\parallel  } \right)$ the total  exciton energy peak:
\begin{equation} 
E\left( {K_\parallel  } \right)
 = E_{gap}  - E_{ex}  + \frac{{\hbar ^2 }}{{2M}}K_\parallel ^2 . 
\end{equation} 
 $E_K $  is the Kane energy ( $E_K $=23eV in GaAs based semiconductors), $m_o $
 the elctron mass and g the spin-degeneracy that, in the present calculation,  it has been taken as a constant value (g = 1). 
                  
Finally, by solving the Lippmann-Schwinger equation\cite{4,24}, we can obtain the  electric field  into explicit form (with the factor $\exp \left[ {iq_x X} \right]$  suppressed):
\begin{eqnarray}
E_\beta  (\omega ;Z) = A_w^{(\beta )} \left[ {e^{ik_w Z}  - g_w (Z)\,\tilde S_{_{ex} }^{(\beta )} (\omega ;q_x )\,\varphi _{ex} (q_x )} \right]
 \nonumber  \\ 
 + B_w^{(\beta )} \left[ {e^{ - iq_x Z}  - g_w (Z)\,\tilde S_{ex}^{(\beta )} (\omega ;q_x )\,\varphi _{ex} ( - q_x )} \right]\ \quad \quad
\end{eqnarray}
 $\varphi _{ex} $ is the Fourier transform of exciton envelope function:
\begin{equation}
\varphi _{ex} ( \pm k_w ) = \int\limits_{ - L_w /2}^{L_{w/2} } {\Psi  (r = 0,Z)\,e^{ \pm iq_x Z} dZ}
\end{equation}
\begin{equation}
\varphi _{ex} ( \pm q_x )  = \frac{{N_{ex} k_{ex}^2 }}{{k_w \left( {k_{ex} ^2  - k_w^2 } \right)}}\sin \left( {\frac{{k_w L_w }}{2}} \right)
\end{equation}
where $k_{ex}  = \pi /L_w $.

We define now the integral function $g_w (Z)$ computed at the slab interfaces ($Z =  \pm L_w /2$):
\begin{equation}
g_w (Z) = \int\limits_{ - L_w /2}^{L_w /2} {dZ''G_w (Z,Z'')\Psi _{ex}^ *  (r = 0,Z'')} 
\end{equation}                 
\begin{equation}
g_w (Z) = \frac{{iN_{ex} k_{ex} ^2 }}{{2q_x \left( {k_{ex} ^2  - q_x^2 } \right)}}\left( {1 - e^{iq_x Z} } \right)
\end{equation}
and, then, the function:
\begin{equation}
\\M_w (\omega ) = \int\limits_{ - L_w /2}^{L_w /2} {dZ} \;\Psi _{ex} (Z)\,g_w (Z)\\
\end{equation}
\begin{eqnarray}
 M_w (\omega ) = \frac{{N_{ex}^2 }}{4}\left[ {L_w \left( {\frac{1}{{q_x^2 }} - \frac{1}{{2\left( {k_{ex} ^2  - q_x^2 } \right)}}} \right) + } \right.  \nonumber \\
 \quad  - \left. {iq_w \left( {\frac{1}{{q_x^2 }} + \frac{1}{{k_{ex} ^2  - q_w^2 }}} \right)^2 \left( {1 - e^{iq_x L_w } } \right)} \right] 
\end{eqnarray}
by means of that we write the new energy function $S_{_{ex} }^{(\beta )} (\omega ;q_x )$:
\begin{equation}
\tilde S_{ex}^\beta   = \frac{{\tilde q_\beta ^2 \;S_o (\omega ) }}{{E_{ex}^2  - \hbar ^2 \omega ^2  - i2\Gamma _{NR} \hbar \omega  + \tilde q_\beta ^2 \;S_o (\omega ) M_w }}
\end{equation}
and obtain the polariton self-energy  $\Sigma _{ex} $ in explicit form:
\begin{equation}
\Sigma _{ex} (\omega ,q_x ) = \tilde q_\beta ^2 \;S_o (\omega )M_w (\omega )
\end{equation} 
\section{UNIAXIAL CRYSTAL SLAB WITH IN-PLANE OPTIC $\hat C$  AXIS}
Let us consider an electromagnetic wave incident from an isotropic medium (vacuum or air)  onto the left surface of an uniaxial multilayer. The Cartesian coordinate system is chosen such that the xz  is the plane of incidence and the  z axis is normal to the reflecting surface. 
The optical response of the uniaxial  multilayer may be characterized  by four reflection amplitudes   $\left( {R_{ss} ,R_{sp} ,R_{ps} ,R_{pp} } \right)$ and four transmission amplitudes  $\left( {T_{ss} ,T_{sp} ,T_{ps} ,T_{pp} } \right)$,  where the second suffix refers to the polarization, S or P, of the reflected and transmitted waves, when the  incident wave is in the polarization state of the first suffix\cite{25,26}.
In order to calculate these amplitudes  we employ the transfer-matrix method  that connect the electromagnetic field amplitudes at both the interfaces of the layers.
Constitutive equations for uniaxial crystal in a Cartesian framework are: 
\begin{equation}
\left\{ {\begin{array}{*{20}c}
   {\left( {q^2 \varepsilon _{xx}  - k_z^2 } \right)E_x  + q^2 \varepsilon _{xy} E_y  + \left( {q^2 \varepsilon _{xz}  + k_\parallel  k_z } \right)E_z  = 0}  \\
   {q^2 \varepsilon _{yx} E_x  + \left( {q^2 \varepsilon _{yy}  - k_\parallel ^2  - k_z^2 } \right)E_y  + q^2 \varepsilon _{yz} E_z  = 0\quad }  \\
   {\left( {q^2 \varepsilon _{zx}  + k_\parallel  k_z } \right)E_x  + q^2 \varepsilon _{zy} E_y  + \left( {q^2 \varepsilon _{zz}  - k_\parallel ^2 } \right)E_z  = 0}  \\
\end{array}} \right.
\end{equation}
where $q = \omega /c$ and  $\varepsilon _{\sigma \nu } $ ($\sigma ,\nu  = x,y,z$) are the components of the Cartesian symmetric dielectric tensor $\varepsilon $, that, for uniaxial crystals with optical axis  $\hat C$ in the reflecting plane (xy) and forming an angle $\alpha $ with the x axis,  becomes:
\begin{equation}
\varepsilon  = \left( {\begin{array}{*{20}c}
   {\varepsilon _ \bot   + s_x^2 \left( {\varepsilon _\parallel   - \varepsilon _ \bot  } \right)} & {s_x s_y \left( {\varepsilon _\parallel   - \varepsilon _ \bot  } \right)} & 0  \\
   {s_x s_y \left( {\varepsilon _\parallel   - \varepsilon _ \bot  } \right)} & {\varepsilon _ \bot   + s_y^2 \left( {\varepsilon _\parallel   - \varepsilon _ \bot  } \right)} & 0  \\
   0 & 0 & {\varepsilon _ \bot  }  \\
\end{array}} \right)
\end{equation}
where $\varepsilon _\parallel  $ and $\varepsilon _ \bot  $ are the dielectric constants, valued in the principal axes framework, that  describe the electromagnetic response for solicitations along the directions parallel and perpendicular to the optical $\hat C$ axis, respectively.  $s_x  = \cos \alpha $ and $s_y  = \sin \alpha $ are the projections of  $\hat C$ on the x and y axes.  We  take nonmagnetic materials for which the magnetic permeability value $\mu  = 1$ is used throughout the whole layered medium.
In order to have nontrivial solutions of Eqs. (B1), the determinant of the coefficient matrix must vanish. This requires the solution of  the fourth degree (in $k_z $) characteristic equation and  the four roots, eigenvalues of the matrix,  are the wave vectors relative to the ordinary $\left( { \pm k_{oz} } \right)$ and extraordinary $\left( {k_{ez}^{\left(  \pm  \right)} } \right)$ two rays: 
\begin{equation}
 k_{oz}  =  \pm \sqrt {q^2 \varepsilon _ \bot   - k_\parallel ^2 } ;
\end{equation}
\begin{equation}
k_{ez}^{\left(  \pm  \right)}  =  \pm \sqrt {\frac{{k_o^2 \varepsilon _\parallel   + k_\parallel ^2 \left[ {s_y^2 \left( {\varepsilon _\parallel   - \varepsilon _ \bot  } \right) - \varepsilon _\parallel  } \right]}}{{\varepsilon _ \bot  }}} 
\end{equation}
where $k_o  = \left( {\omega /c} \right)\sqrt {\varepsilon _ \bot  } $, $k_\parallel   = k_x  = \left( {\omega /c} \right)\sqrt {\varepsilon _1 } \sin \theta _1 $.  $\theta _1 $ is the angle of incidence and $\varepsilon _1  = 1$.  Notice that the expressions in Eqs. (B3) and (B4) are all independents of the polarization state of the incident wave.
  The corresponding twelve solutions of Eq. (B1), are $E_{o\nu }^ \pm  e^{ \pm ik_{oz} z} $ $\left( {\nu  = x,y,z} \right)$ for the ordinary ray and  $E_{e\nu }^ \pm  e^{ik_{ez}^ \pm  z} $   for the extraordinary ray:
\begin{eqnarray}
E_{ox}^{\left(  \pm  \right)}  =  \mp s_y k_{oz} \quad\quad E_{ex}^{\left(  \pm  \right)}  = s_x k_{oz}^2 \quad  \\  \nonumber
E_{oy}^{\left(  \pm  \right)}  =  \pm s_x k_{oz} \quad\quad E_{ey}^{\left(  \pm  \right)}  = s_y k_o^2 \quad  \\ \nonumber
E_{oz}^{\left(  \pm  \right)}  = s_y k_\parallel \quad \quad E_{ez}^{\left(  \pm  \right)}  =  - s_x k_{ez}^{\left(  \pm  \right)} k_\parallel .
\end{eqnarray}
Finally, each   $E_\nu  $ component    $\left( {\nu  = x,y,z} \right)$ of the electric field is obtained as a linear combination of the corresponding four partial waves:
\begin{widetext}
\begin{equation}
E_\nu   = \left( {A_o E_{o\nu }^ +  e^{ik_{oz} z}  + B_o E_{o\nu }^ -  e^{ - ik_{oz} z}  + A_e E_{e\nu }^ +  e^{ik_{ez}^ +  z}  + B_e E_{e\nu }^ -  e^{ik_{ez}^ -  z} } \right)e^{ik_\parallel  x} .
\end{equation}
\end{widetext}
The magnetic field components are then obtained by the Maxwell equation $
\bf{\nabla }  \times {\bf{E}} = i\omega \mu _0 \mu {\bf{H}}
$.
The four coefficients $A_o ,B_o ,A_e $ and $B_e $ are obtained, for each layer of heterostructure, by imposing the continuity of the tangential components, of both the electric and magnetic fields, at the boundaries of each layer together to the following propagation relation:
\begin{equation}
C_j \left( \begin{array}{l}
 E_x \left( {z_j } \right) \\ 
 E_y \left( {z_j } \right) \\ 
 H_x \left( {z_j } \right) \\ 
 H_y \left( {z_j } \right) \\ 
 \end{array} \right) = \left( \begin{array}{l}
 E_x \left( {z_j  + L_j } \right) \\ 
 E_y \left( {z_j  + L_j } \right) \\ 
 H_x \left( {z_j  + L_j } \right) \\ 
 H_y \left( {z_j  + L_j } \right) \\ 
 \end{array} \right)
\end{equation}
where $z_j $ and $L_j $ are the coordinate of the left interface and the thickness of the layer respectively.  The transfer matrix $C_j $  is easily building up, for each j-layer, as:
\begin{equation}
C_j  = M_j U_j (L_j )M_j^{ - 1} 
\end{equation}
where:
\begin{equation}
M_j  = \left( {\begin{array}{*{20}c}
   {E_{ox}^ +  } & {E_{ox}^ -  } & {E_{ex}^ +  } & {E_{ex}^ -  }  \\
   {E_{oy}^ +  } & {E_{oy}^ -  } & {E_{ey}^ +  } & {E_{ey}^ -  }  \\
   {H_{ox}^ +  } & {H_{ox}^ -  } & {H_{ex}^ +  } & {H_{ex}^ -  }  \\
   {H_{oy}^ +  } & {H_{oy}^ -  } & {H_{ey}^ +  } & {H_{ey}^ -  }  \\
\end{array}} \right)
\end{equation}
\begin{equation}
U(L_j ) = \left( {\begin{array}{*{20}c}
   {e^{ik_{joz} L_j } } & {} & {} & {}  \\
   {} & {e^{ - ik_{joz} L_j } } & {} & {}  \\
   {} & {} & {e^{ik_{jez}^ +  L_j } } & {}  \\
   {} & {} & {} & {e^{ - ik_{jez}^ +  L_j } }  \\
\end{array}} \right)
\end{equation}

The transfer Matrix C of the whole n-layer stack is: $C = C_n C_{n - 1} .....C_1 $.
Finally we can write the following relations between the incoming and outgoing electromagnetic fields for S polarized incident wave:
\begin{equation}
C\left( {\begin{array}{*{20}c}
   {\cos \theta _1 R_{sp} }  \\
   {1 + R_{ss} }  \\
   { - k_1 \cos \theta _1 \left( {1 - R_{ss} } \right)}  \\
   { - k_1 R_{sp} }  \\
\end{array}} \right) = \left( {\begin{array}{*{20}c}
   {\cos \theta _2 T_{sp} }  \\
   {T_{ss} }  \\
   { - k_2 \cos \theta _2 T_{ss} }  \\
   {k_2 T_{sp} }  \\
\end{array}} \right)
\end{equation}
and for P polarized incident wave:
\begin{equation}
C\left( {\begin{array}{*{20}c}
   {\cos \theta _1 \left( {1 + R_{pp} } \right)}  \\
   {R_{ps} }  \\
   {k_1 \cos \theta _1 R_{ps} }  \\
   {k_1 \left( {1 - R_{pp} } \right)}  \\
\end{array}} \right) = \left( {\begin{array}{*{20}c}
   {\cos \theta _2 T_{pp} }  \\
   {T_{ps} }  \\
   { - k_2 \cos \theta _2 T_{ps} }  \\
   {k_2 T_{pp} }  \\
\end{array}} \right)
\end{equation}
where  $k_1  = q\sqrt {\varepsilon _1 } $, $k_2  = q\sqrt {\varepsilon _2 } $, $\theta _1 $ and $\theta _2 $ angles of incidence and transmission respectively.  The refraction (R) and transmission (T) amplitudes are then computed as the ratio of the  intensities  of the  electric field, of the reflected or transmitted waves, to the incident electric field intensity.  
\\
\section*{Acknowledgments}
The Authors are indebted with Dr. L. Pilozzi for the critical reading of the manuscript.
\end{document}